\documentclass[10pt, conference, compsocconf]{IEEEtran}
\ifCLASSINFOpdf
 \usepackage[pdftex]{graphicx}
\else
\fi

 \usepackage{url}
 \usepackage{flushend}

\begin{document}
\title{Characterizing the Cloud's Outbound Network Latency: An Experimental and Modeling Study %based on Google App Engine
\thanks{This work is supported in part by Chilean National Commission for Scientific and Technological Research (CONICYT, Chile) under Grant
FONDECYT Iniciaci{\'o}n 11180905.}
}

\author{\IEEEauthorblockN{Zheng Li}
\IEEEauthorblockA{\textit{Department of Computer Science} \\
\textit{University of Concepci{\'o}n}\\
Concepci{\'o}n, Chile \\
ORCID: 0000-0002-9704-7651}
\and
\IEEEauthorblockN{Francisco Millar-Bilbao}
\IEEEauthorblockA{\textit{Department of Computer Science} \\
\textit{University of Concepci{\'o}n}\\
Concepci{\'o}n, Chile \\
Email: frmillar@udec.cl}
}

\maketitle

\begin{abstract}
Cloud latency has critical influences on the success of cloud applications. Therefore, characterizing cloud network performance is crucial for analyzing and satisfying different latency requirements. By focusing on the cloud's outbound network latency, this case study on Google App Engine confirms the necessity of optimizing application deployment. More importantly, our modeling effort has established a divide-and-conquer framework to address the complexity in understanding and investigating the cloud latency. 

\end{abstract}

\begin{IEEEkeywords}
cloud application; data transmission; geographical location; Internet topology; outbound network latency;

\end{IEEEkeywords}

% \fi
\IEEEpeerreviewmaketitle

\section{Introduction}
Given the distributed nature of cloud computing, many cloud features are subject to uncertainty to some extent \cite{Mezni_Aridhi_2018}. In particular, the outbound network latency of the cloud can be impacted by various subtleties that are difficult to identify \cite{Strom_Frank_2020}, not to mention that the users' lack of infrastructural controllability further increases the challenges in understanding the cloud network performance. For example, the IP addresses of Google cloud are generally registered under Google's headquarters in Mountain View, California, even if the actual servers are located in different countries; consequently, it becomes impossible to use traditional benchmark (e.g., ping and traceroute) to quickly reveal the cloud latency in this case.

Considering that it is crucial to select suitable cloud environments to match the latency requirements of different applications \cite{Strom_Frank_2020}, we decided to develop a generic methodology to characterize the cloud network performance. The methodology includes an iterative process of experimental investigation and modeling investigation. On one hand, the modeling work can help interpret the existing experimental results and guide upcoming experiments. On the other hand, by fitting the experimental data to the models, we will be able to better understand the nature of cloud latency and then estimate/optimize the performance of cloud applications.

Our current focus is on profiling the cloud's outbound network latency, and this paper reports our ongoing case study on Google App Engine (GAE). In addition to giving a holistic view of the outbound latency with respect to GAE's 17 regions visited from six continents, this study's major contribution is the innovation in our modeling investigation: (1) The initially established models have suggested a divide-and-conquer approach to addressing the complexity in understanding cloud latency. (2) We developed a new metric of data transmission speed (measured by $\mathit{Byte\cdot meter/sec}$) to involve transmission distance that is missing in the traditional metrics like bandwidth and throughput. 

\section{Experimental Investigation}
\subsection{Setting Up the Testbed}
To test the cloud's outbound network latency, we employ the downloading scenario that represents one of the elementary activities of cloud applications \cite{Li_Tesfatsion_2017}, i.e.~a client downloads data blobs from a server in the cloud. Correspondingly, we follow the client-server model to architect and build up the testbed. In specific, \textit{(i)} we deploy one data server to each of GAE's available regions;\footnote{The detailed information about the 17 data servers are shared online at \url{https://bit.ly/35pLaKt}} \textit{(ii)} after empirically deciding the optimal data size to be about 10 MB for our testbed, we choose a video file (11.2 MB) as the data to be downloaded;\footnote{An example data server is at \url{https://sydney-server-mt.appspot.com/HQ}} \textit{(iii)}  we employ Amazon's micro EC2 instances to play the client role from six continents. Note that the investigation object of this study is still GAE rather than Amazon EC2.

\subsection{Experimental Implementation and Results}
After setting up the testbed, we exhaustively measured the file downloading latency with each of the $6 \times 17$ client-server pairs. In particular, to make the  intra-continent latency measurements more comparable,  we have tried to select client locations (i.e.~Amazon's cloud regions) where the host cities also have GAE regions (except for Cape Town because GAE does not have a region in Africa). 

By sorting the client and server locations according to their air distances\footnote{The air distance between two cities are obtained through \url{ https://www.distancefromto.net/}} to a reference city (i.e.~Ashburn where an edge spot of Google Cloud exists), we visualize the measurement results into a heat map.\footnote{Heat map of GAE's outbound network latency (measured in $\mathit{ms}$) is shared onine at \url{https://bit.ly/315TRJw}} Benefiting from the visualization, we summarize a set of typical observations:

\begin{itemize}
\item In general, the cloud's outbound data transfer has lower latency within the intra-continent range than that within the inter-continent range. 
\item By particularly focusing on North America and Europe, the intra-continent distance seems to have a roughly linear proportional impact on the cloud's outbound network latency.
\item In contrast, the cloud's outbound network latency does not show any clear correlation with the geographical distance to the  inter-continent clients. 
\item Surprisingly, the outbound latency does not seem to be symmetrical in terms of mutual network traffic between two cloud regions. 
\end{itemize}

% Note that IEEE typically puts floats only at the top, even when this
% results in a large percentage of a column being occupied by floats.

\section{Modeling Investigation}
Modeling is a powerful approach to understanding real-world objects and processes that are difficult to observe directly \cite{Basmadjian_2019}.
Since cloud data transmission relies on the Internet, we refer to the three-layer hierarchical representation of Internet \cite{Hinton_Baliga_2011} to interpret cloud's outbound networking.

Recall that the client nodes in our testbed are also deployed in the cloud data centers. Then, the measured cloud outbound network latency in our study is only related to the core network and the metro/edge network. Thus, we model the outbound network latency $L$ to be the overall time consumption of data transmission within two (Google's and Amazon's) cloud data centers (denoted by $2D/B_\mathit{ds}$), the data traverse across two corresponding regions' core networks (denoted by $2D/B_c$), and the one-way traverse via the metro/edge network (denoted by $L_m$), as defined in Eq.~(\ref{eq:latency}).

\begin{footnotesize}
\begin{equation}
\label{eq:latency}
%  L = 2L_\mathit{DS} + 2L_C + 2L_\mathit{CM} + L_M
L = 2\frac{D}{B_\mathit{ds}} + 2\frac{D}{B_c} + L_m
\end{equation}
\end{footnotesize}%
where $D$ represents the data size, $B_\mathit{ds}$ indicates the network bandwidth within the data center, and $B_c$ indicates the bandwidth of the core network. %the time consumption $2L_\mathit{CM}$ for data exchange between the core network and the metro/edge network, 
 
Benefiting from the modeling, we can investigate the cloud's outbound network latency through a divide-and-conquer approach. In fact, we have conservatively estimated $B_\mathit{ds}$ to be 1.25GB/sec, according to Google's dedicated network technologies for server connections in the data centers \cite{Vahdat_2015}. At the time of writing, we are empirically investigating $B_c$ within different cloud regions. As for $L_m$, ideally, it could be as trivial as the overhead of data exchange between neighboring core nodes (e.g., the Amazon's and Google's regions both in Frankfurt). However, in a generic sense, $L_m$ needs to be further specified by taking into account data transmission through land and submarine cables. Moreover, we are concerned with the practical speed (measured by $\mathit{Byte\cdot meter/sec}$) instead of the cable bandwidth in this case, because it is known that distance also matters in the wide-range data transmission.

Given the observations in our experimental investigation, we further model the metro/edge networking delay $L_m$ as: 

\begin{footnotesize}
\begin{equation}
\label{eq:MetroLatency}
  L_m = \frac{D\cdot \mathit{I_{lan}}}{S_\mathit{lan}} + \frac{D\cdot \mathit{I_{sub}}}{S_\mathit{sub}} + L_r\propto N
\end{equation}
\end{footnotesize}%
where $S_\mathit{lan}$ and $S_\mathit{sub}$ represent the inland and submarine data transmission speeds respectively, while  $\mathit{I_{lan}}$ and $\mathit{I_{sub}}$ indicate the distances of data transmission via the inland and submarine cables respectively. In particular, we have used geographical distance to estimate the inland data transmission distance between two cities in the same continent.

Note that, in addition to the inland latency $D\cdot \mathit{I_{lan}}/S_\mathit{lan}$ and the submarine latency $D\cdot \mathit{I_{sub}}/S_\mathit{sub}$, we also highlight an extra transmission overhead $L_r$ that is proportional to the amounts $N$ of inland-submarine data relaying, so as to reflect different topologies of the inter-continent networking. %For example, from the Tokyo-side client's perspective, GAE's South Carolina region and London region have nearly the same outbound network latency. Considering the relatively poor connectivity of submarine cables in this case,\footnote{\url{https://www.submarinecablemap.com/}} the data relaying must have dominated the causes of transmission delay. 

Overall, by fitting the experimental data to this modeling work, we are able to build regression-based formulas to facilitate performance estimation and optimization of cloud applications.  

\section{Conclusion}
It is evident that reducing latency has tremendous effects on the success of cloud applications. %Therefore, before releasing cloud applications, it will be crucial and valuable to characterize the cloud environment's network performance to optimize the application deployment. 
By focusing on the outbound network latency, this case study not only reveals useful information for applications to be deployed on GAE, but also builds a spiral path to guide generic investigations into the cloud latency.

%\section*{Acknowledgment}
%The funding information is removed to satisfy the requirement of double-blind reviewing.

%\newcommand{\BIBdecl}{\setlength{\itemsep}{0.5 em}}
\newcommand{\BIBdecl}{\setlength{\itemsep}{1 ex}}
\bibliographystyle{IEEEtran}
{\footnotesize
\bibliography{Cloud20RefMini}
}

\end{document}